\documentclass[twocolumn,showpacs,preprintnumbers,amsmath,amssymb]{revtex4}

\usepackage{graphicx}
\usepackage{dcolumn}
\usepackage{bm}

\begin{document}

\title{An intrinsic peak-dip-hump and strong coupling effects in
\\ $Bi_{2}Sr_{2}CaCu_{2}O_{8+\delta}$ ARPES data near $(\pi,0)$}

\author{A. D. Gromko$^{1}$, Y.-D. Chuang$^{1,2}$, A. V.
Fedorov$^{1,2}$, Y. Aiura$^{3}$, Y. Yamaguchi$^{3}$, K. Oka$^{3}$,
Yoichi Ando$^{4}$, D. S. Dessau$^{1}$}

\affiliation{$^{1}$Department of Physics, University of Colorado,
Boulder, Colorado 80309-0390}

\affiliation{$^{2}$Advanced Light Source, Lawrence Berkeley
National Laboratory, Berkeley, CA 94720}

\affiliation{$^{3}$
AIST, AIST Tsukuba Central 2, 1-1-1
Umezono, Tsukuba, Ibaraki 305-8568, Japan}

\affiliation{$^{4}$
CRIEPI, 2-11-1 Iwato-Kita, Komae, Tokyo 201-8511,
Japan}

\date{\today}

\begin{abstract}

A well-known peak-dip-hump structure exists near $(\pi, 0)$ in
superconducting state ARPES spectra of
$Bi_{2}Sr_{2}CaCu_{2}O_{8+\delta}$ (Bi2212). Here we report
results on optimal and overdoped Bi2212 samples indicating the
traditional peak-dip-hump structure observed near $(\pi,0)$ is
largely due to bilayer splitting. However a separate, much weaker
peak-dip hump (PDH) structure distinct from bilayer splitting can
be detected near $(\pi, 0)$. This new PDH structure is consistent
with electronic coupling to the magnetic resonance mode in Bi2212.
Both the dispersion and line shape signatures indicate strong
coupling to this mode.

\end{abstract}


\maketitle

The anomalies observed in the normal and superconducting-state
electronic structure of the cuprates verify the complexity and
richness of High-Temperature Superconductivity
(HTS)~\cite{Orenstein:2000}. Angle-resolved photoemission
spectroscopy (ARPES) has proven to be an invaluable tool for
advancing our understanding of HTS due to the energy and
momentum-dependent information it provides. One of the most
dissected features of ARPES spectra from the cuprate
$Bi_{2}Sr_{2}CaCu_{2}O_{8+\delta}$ (Bi2212) has been the so-called
peak-dip-hump (PDH) structure found in the superconducting state
energy distribution curves (EDCs) near the $(\pi, 0)$ point of the
Brillouin zone.  This structure consists of a sharp, low binding
energy ($\sim 30 meV$) peak, a broad high binding energy ($\sim
100 meV$) hump, and a spectral dip between
them~\cite{Dessau:1991,Norman:1997,Campuzano:1999,Kaminski:2001}.
Similar structures are observed in tunnelling spectra from both
HTSs~\cite{Renner:1995,Zasadinski:2001} and strongly-coupled
conventional superconductors, where in the latter the PDH has been
shown to match the phonon density of states measured in neutron
scattering experiments~\cite{McMillan:1969}. The successful
identification of electron-phonon coupling as the mechanism for
conventional superconductivity increases the importance of
understanding the PDH structure in HTSs.

In light of this, many well-known theories such as the marginal
Fermi Liquid~\cite{Varma:1989} and resonating valence bond
interlayer tunnelling theories~\cite{Anderson:1997} have attempted
to incorporate the PDH structure. More recently, it has been
widely discussed in terms of coupling to some boson, with
particular attention paid to the magnetic resonance mode observed
in inelastic neutron scattering (INS)
experiments~\cite{Bourges:1998}. One proposed connection between
the INS and ARPES results relates the energy of the magnetic mode
to the energy position of the dip~\cite{Norman:1997,
Campuzano:1999, Kaminski:2001}. Alternatively, other reports
connect the weight of the sharp peak to a superfluid condensate
fraction~\cite{FengandDing:2000}. All of these ideas model a
single, broad normal state peak replaced by the PDH structure in
the superconducting state. However, recent advances in ARPES
normal state data, namely the detection of bilayer-band splitting
(coupling between $CuO_{2}$ planes within a unit
cell)~\cite{Chuang:2000,Feng:2001}, show that the normal state
spectra display both bonding (B) and antibonding (A) band features
near $(\pi,0)$ instead of a single peak. With this new knowledge,
it is natural to question the effect of bilayer splitting on the
superconducting state spectra and the
PDH~\cite{Chuang:2000,Feng:2001,Kordyuk:preprint}.  Here we argue
that the classical PDH structure observed near $(\pi,0)$ in Bi2212
is in fact an artifact of bilayer splitting. In addition, we
report the first observations of a new PDH near $(\pi,0)$
contained within a single band. This new peak-dip-hump feature is
argued to be intrinsic and of a lower energy scale and strength
than the classical PDH. The strength of the dispersion kink
associated with the new peak-dip-hump structure suggests strong
electronic coupling.

The data presented here were taken at the Stanford Synchrotron
Radiation Laboratory (SSRL), Stanford, and at the Advanced Light
Source (ALS), Berkeley. At both facilities we used Scienta 200mm
electron spectrometers, allowing the simultaneous collection of
data along a $\sim 14^{\circ}$ angular slice with $0.08^{\circ}$
resolution along the slice. The beamline and analyzer slits were
adjusted to achieve an experimental energy resolution of 12meV, as
determined by the 10-90$\%$ width of a gold Fermi edge. The photon
energy was tuned to 20eV, with the polarization along the
$(0,0)-(\pi,0)$ direction. The analyzer slit direction was
parallel to the $(0,0)-(\pi,0)$ direction for all cuts (see inset,
figure 1) . We present data from Bi2212 samples at four doping
levels: heavily overdoped OD58 $(T_{c}=58K, \Delta T_{c}=3K)$ and
OD71 $(T_{c}=71K, \Delta T_{c}=4K)$, optimally doped OP91
$(T_{c}=91K, \Delta T_{c}=2K)$, and lightly underdoped UD85
$(T_{c}=85K, \Delta T_{c}=6K)$.

Figure 1(a)-(l) shows a sampling of raw data for OP91 and OD71
both above and below $T_{c}$, presented as false-color scale
intensity plots. In general, two features are observed in each
panel, which are lableled as the antibonding (A) and bonding (B)
bands due to bilayer splitting.  Superstructure (SS) bands due to
the lattice mismatch of the $BiO$ and $CuO_{2}$ planes are
occasionally seen as well. While bilayer splitting is well
accepted for the overdoped regime of
Bi2212~\cite{Chuang:2000,Feng:2001,Kordyuk:preprint}, the data in
panels (k) and (l) are probably the clearest data to date
demonstrating the persistence of bilayer splitting to lower doping
levels where the instrinsic line broadening makes the
deconvolution into separate features more difficult.

We now examine how bilayer splitting manifests itself in the EDC
line shape along the well-studied $(0,0)-(\pi,0)$ symmetry line.
Figures 1(m) and (n) show EDCs in both the normal and
superconducting states along $(0,0)-(\pi,0)$. These curves are
taken as vertical (energy) slices from the center cuts for OD71,
and from cuts similar to and including those in panels (k) and (l)
for OP91. The EDCs at both temperatures show two dominant spectral
features, a low energy ($\sim 30meV$) sharp peak and a higher
energy ($\sim 100meV$) broad hump. From panels (a)-(j), we see
that the energy scales of the two features directly correspond to
the A and B bands for sample OD71. Similar agreement is observed
for the full data set on OP91, not shown here. From this agreement
we deem the EDC line shape along $(0,0)-(\pi,0)$ to be a direct
result of bilayer splitting rather than the signature of
self-energy effects. In fact, recent studies indicate the
intensities of the two EDC peaks vary independently with photon
energy~\cite{Kordyuk:preprint,Chuang:preprint}. Our
temperature-dependent data in figure 1 is complimentary to these
studies, since it shows the "classic" PDH structure is present in
both the normal and superconducting states in both overdoped and
optimally doped samples. Although at present the severe spectral
broadening found in underdoped samples~\cite{DingandLoeser:1996}
disallows accurate temperature-dependent measurements of bilayer
splitting, the aforementioned photon energy-dependent reports
indicate bilayer splitting is present in the normal state of
underdoped Bi2212 samples~\cite{Chuang:preprint}.

Although our data strongly indicates that the standard PDH
structure at $(\pi,0)$ is due simply to bilayer splitting, there
is a new PDH structure appearing in the superconducting state data
near $(\pi,0)$ associated with a new energy scale $E_{kink}$. Kink
effects have been discussed recently in ARPES studies of
HTS's~\cite{Lanzara:2001,Johnson:2001,Gromko:preprint,Kaminski:2001},
mostly connected with the nodal region and obtained from momentum
distribution curve (MDC) data, as illustrated for UD85 in figure
2(a). We present underdoped data since in these samples the nodal
kink or self-energy effects are
strongest~\cite{Lanzara:2001,Johnson:2001,Gromko:preprint}. The
temperature dependence of the nodal kink is very weak, as
demonstrated by the similarity of the normal state (black dots)
and superconducting state (red dots) MDC-derived dispersions.
Figure 1 shows that the data near $(\pi,0)$ (for example panels
(e) and (j)) has a more significant change in dispersion than the
nodal data~\cite{Gromko:preprint}.

This can be more clearly seen in figure 2(b), which shows a blowup
up of $k_{y}=1.0\pi$ superconducting state data from sample OD58.
On the graph we have traced a segment of the dispersion of the A
and B bands, for which we have made use of both EDCs (black and
blue dots) and MDCs (red dots)~\cite{EDCandMDC}. The possibility
that the MDC and EDC peak positions do not match in the presence
of self-energy effects has been pointed out in the
literature~\cite{Kaminski:2001,Gromko:preprintold} and is
especially clear in this data near the kink energy scale of $40
meV$, shown by the horizontal blue line. While the kink energy
scale shows up in the MDC-derived dispersion (red), it is even
more clear in the EDC dispersion (blue), which asymptotically
approaches the kink energy scale. In this instance, the
disagreeing portions of the MDC and EDC dispersion represent the
splitting of the B band dispersion into two branches. The low
energy EDC dispersion tracks the renormalized part of the B band
dispersion ($B^{\prime}$), while at energies above $E_{kink}$ the
MDC dispersion tracks the unrenormalized part
($B^{\prime\prime}$). The EDC dispersion also roughly tracks
$B^{\prime\prime}$ at binding energies below $E_{kink}$, however
as shown in figure 2(c) these features are broad and hence the
dispersion is not shown. Previously, from MDC data only, we argued
that the detailed temperature, momentum, and doping dependence of
the kink scale at $(\pi,0)$ indicate that this kink is clearly
different from and stronger than the nodal kink, and that it is
likely a result of electronic coupling to the magnetic resonance
mode observed in inelastic neutron
scattering~\cite{Gromko:preprint}. Two branches can be understood
in terms of coupling to a bosonic
mode~\cite{Norman:1997,Scalapino:1969}, as only virtual
excitations can be excited below $E_{kink}$ while real ones which
damp the system can be excited above $E_{kink}$. The peak
intensities (related to the imaginary part of the self-energy) are
also consistent with this, showing a large increase in spectral
weight below $E_{kink}$.

We now show the new PDH in the EDCs due to the kink effect. At
$k_{x}=0.1\pi$ (figure 2(c)), the normal state EDC basically shows
one broad feature cut by the Fermi function, which we now know to
be a superimposition of the A and B bands. As the sample goes
superconducting, the EDC line shape is transformed into two sharp
low binding energy peaks and a broader high binding energy peak.
If we look at the intensity plots, these features are easily
understood. The peak closest to $E_{F}$ corresponds to the A band
(black dots). The second sharp peak corresponds to the
quasiparticle pole ($B^{\prime}$) of the B band (blue dots). The
broad hump is the higher binding energy branch
($B^{\prime\prime}$) of the B band (red dots). The fact that the B
band dispersion splits into two branches, one below ($B^{\prime}$)
and one above ($B^{\prime\prime}$) the kink energy (blue solid
line, panel (b)), is a strong coupling effect. For $\vec k$ values
close to where $B^{\prime}$ meets the asymptote, the $B^{\prime}$
and $B^{\prime\prime}$ EDC features (panel (c)) comprise a true
PDH structure, with the $B^{\prime}$ peak asymptotically defining
the kink energy (blue dashed line). The A band feature does not
develop a PDH but only sharpens upon cooling since even in the
normal state it is below the kink energy for all $\vec k$ values
in the slice. If we now move to the $\vec k$-value where the band
crosses $E_{F}$ in the normal state $(k_{x}=0.19\pi)$, we see only
the quasiparticle peak $B^{\prime}$ (figure 2(f)).

Three spectral features similar to what we observe at
$k_{x}=0.1\pi$ have been reported by Feng et al.~\cite{Feng:2001}.
They argued that upon going superconducting, the A and B bands
both moved to lower energies and obtained more than a factor of
four reduction in their energy splitting, as if the intralayer
coupling $t_{\perp}$ was reduced in the superconducting state.
Although they did not report any kink effects or any specific
energy scale, they guessed that each of the low energy A and B
bands should form their own PDH structure.  This implies four
peaks total (two peaks and two humps), although they were not able
to resolve two hump structures.  With the clear observation of the
kink and its associated energy scale in our new work, we are able
to understand their observations.  First, since the normal state A
band dispersion is below $E_{kink}$ throughout this entire cut, it
never develops a hump structure, while the B band develops its PDH
only near where the normal state B band dispersion crosses
$E_{kink}$.  Along other cuts away from $(\pi,0)$ the A band will
disperse across $E_{kink}$ hence developing a PDH structure,
although the kink or coupling strength weakens away from $(\pi,0)$
so this effect may never be observable.  The $B^{\prime}$ peak
does not extend to the $(\pi,0)$ point as sketched by Feng et al.,
but rapidly dies away as spectral weight is transferred to the
$B^{\prime\prime}$ hump. This limited momentum span is due to
finite coupling, an estimate of which will be presented below. As
coupling grows, more spectral weight will be transferred from
$B^{\prime\prime}$ to $B^{\prime}$, enabling the $B^{\prime}$
dispersion to be visible over a larger momentum range.

If we now move to $k_{x}=0$ on figure 2(b) and examine the EDC
temperature dependence (figure 2(e)), we see that the coupling is
not large enough for the $B^{\prime}$ dispersion to extend to
$(\pi,0)$. Consistent with the analysis of figure 1, we see the
low energy peak of the A band $\sim 20meV$ and broad hump
corresponding to the $B^{\prime\prime}$ branch at $\sim 100meV$,
with no sign of the $B^{\prime}$ peak. The only signature of the
kink energy scale (blue dashed line) is an increased coherence of
the A band peak. Previous
ARPES~\cite{Norman:1997,Campuzano:1999,Kaminski:2001} and
tunnelling~\cite{Zasadinski:2001} studies have attempted to relate
the INS resonance energy $\Delta +\omega_{R}$ to the energy of the
dip in the peak-dip-hump structure at $(\pi,0)$. Figures 2(c) and
(e) demonstrate that the dip energy scale (red dashed line) in
general does not match the kink energy scale (blue dashed line).

To produce a strong kink in the B band dispersion and the
accompanying PDH in the EDC lineshape, a fairly large coupling
strength (with a dimensionless coupling constant $\lambda$ of
order of 1 or more) is required. The mass renormalization
associated with the electronic coupling is related to the
quasiparticle velocity ($\frac{1}{\hbar} \frac {dE}{d\vec k}$),
which is decreased by the factor $1+\lambda$ below the energy
scale of the kink. For example, given a coupling strength of
$\lambda=1$, the low energy dispersion will be half as steep as it
is in the absence of the coupling. As the dispersion is directly
measurable in ARPES, the coupling strengths are also in principle
directly measurable, assuming that a "non-interacting" dispersion
can be determined. For data along the nodal direction
$(0,0)-(\pi,\pi)$ (figure 2(a)), where the superconducting gap
goes to zero, the coupling strength has been estimated in several
different ways~\cite{Lanzara:2001,Johnson:2001}. Only a single
energy band is present since bilayer splitting goes to zero
here~\cite{Chuang:2000,Feng:2001}. One method for estimating
$\lambda$ is to examine the difference between the superconducting
(red dots) and normal state (black dots) dispersions at low
energy, which implicitly assumes that the coupling is not present
in the normal state. From figure 2(a) we estimate $\lambda \sim
0.1$ using this method, consistent with the observation that the
normal and superconducting state MDC dispersions are very
similar~\cite{Lanzara:2001}. An alternative way is to consider a
"non-interacting" dispersion obtained by extrapolating the high
energy dispersion to zero energy at $k_{F}$ (black line in figure
2(a)).  This method gives a coupling strength of $\lambda \sim
0.7$ for both the normal and superconducting states.

For the data of figure 2(b) at $(\pi,0)$, the superconducting gap
complicates the issue of accurately extracting $\lambda$. To
estimate the "non-interacting" dispersion we might start with that
expected from the Bardeen-Cooper-Schrieffer (BCS) theory.  In BCS,
the superconducting state dispersion is $E(\vec k)=\surd
(\epsilon(\vec k)^{2}+\Delta(\vec k)^{2})$, where $\epsilon(\vec
k)$ is the normal state dispersion and $\Delta(\vec k)$ is the
gap. Here we use $\Delta=18 meV$ defined as the minimum energy of
the EDC peak. For $\epsilon(\vec k)$ we use the MDC dispersion
derived from the normal state data, as this data does not display
any kinks (figure 1). The resulting $\epsilon(\vec k)$ is
indicated by the black line in figure 2(b). It is clear that the
measured dispersion (either EDC or MDC) deviates significantly
from the BCS prediction. This difference points to significant
interaction effects. Focussing on the EDC dispersion near the gap
edge, we see that it is significantly flatter than the calculated
BCS dispersion, as if a renormalized $\epsilon(\vec k)$ was
gapped. Parameterizing the low energy portion ($-18$ to $-30 meV$)
of the superconducting data by $E(\vec k)=\surd ( \frac
{\epsilon(\vec k)}{1+\lambda}^{2}+\Delta(\vec k)^{2})$ gives the
best agreement with $\lambda=1.05$. This analysis ignores the
information contained in the strength of the kink. Preliminary
analysis of the $(\pi,0)$ kink strength in this data within the
spin-fluctuation model also indicates a lambda close to
1~\cite{Abanov:pcomm}. It also assumes that there is no coupling
in the normal state, the validity of which is not clear at this
time and requires further study. Instead, the lack of a kink in
the normal state may be due to the energy smearing of the magnetic
excitations~\cite{Bourges:2000}. Regardless of the details, a
simple comparison of the $(\pi, 0)$ data to the nodal data in
figure 2(a) indicates that the coupling effects at $(\pi,0)$ are
much stronger than they are along the node, which had previously
been argued to show strong coupling~\cite{Lanzara:2001}.
Consistent with this, the normal and superconducting EDCs along
the nodal direction show little observable PDH effect, as can also
be seen by the absence of two dispersion branches near $E_{kink}$
in figure 2(a). This result conflicts with earlier reports of a
PDH along the nodal direction~\cite{Lanzara:2001}. These points
support the viewpoint that the electronic interactions near
$(\pi,0)$ and along $(0,0)-(\pi,0)$ are distinctly different in
nature. We also note that this work directly contradicts the
theoretical arguments of Kee et al.~\cite{Kee:preprint}, who argue
that the electron-resonance mode coupling should have a maximal
$\lambda$ of order 0.05, i.e. the resonance mode should be highly
irrelevant to the superconductivity. Theoretical arguments
favoring a sizeable $\lambda$ are contained in
ref~\cite{Abanov:preprint}.

We acknowledge beamline support from X.J. Zhou, P. Bogdanov, D.H.
Liu, Z. Hussain, and Z.-X. Shen, and helpful discussions with A.
Chubukov, C. Kendziora, D. Pines, D. Scalapino, and J. Schmallian.
We gratefully acknowledge the help of R. Goldfarb at NIST for the
use of the SQUID magnetometer. This work was supported by the NSF
Career-DMR-9985492 and the DOE DE-FG03-00ER45809.  ALS and SSRL
are operated by the DOE, Office of Basic Energy Sciences.

%
%
%
\begin{figure*}
\caption{\label{figure1} (a)-(j) ARPES data from OD71 at
temperatures above (top row) and below (bottom row) $T_{c}$. The
angular cuts are parallel to the $(\pi,\pi)-(\pi,0)-(\pi,-\pi)$
symmetry line (blue bar, panels (m,n) inset). The $k_{y}$ location
of each cut is labelled on each panel. (k),(l) ARPES data from
OP91 above and below $T_{c}$ for cuts at $k_{y}=0.8\pi$. (m),(n)
EDC data along the $(0,0)-(\pi,0)$ symmetry line ($k_{x}=0$) for
both the normal (red curves) and superconducting (blue curves)
states. The inset shows the EDC locations in the 2D Brillouin zone
as open (OP91) and closed (OD71) circles.}
\end{figure*}
%
%
%

%
%
%
\begin{figure*}
\caption{\label{figure2} (a),(b) Superconducting state ARPES data
from UD85 along $(0,0)-(\pi,\pi)$ and from OD58 near $(\pi,0)$
(blue bars, panel (d)). On panel (a), the normal state MDC-derived
dispersion (black dots) is plotted in addition to the
superconducting state dispersion (red dots). On panel (b) the A
and B band MDC and EDC dispersions are plotted as discussed in the
text. (c),(e),(f) EDCs in the normal (red) and superconducting
(blue) states extracted from OD58 at the labelled momentum
locations, also shown as dashed lines on panel (b).}
\end{figure*}
%
%
%


\begin{references}
\bibitem{EDCandMDC} The use of both EDCs and MDCs to determine dispersion has
become well accepted, with the various methods most useful in
opposite regimes.  In particular, the EDCs are most useful where
the dispersion is very flat, e.g. for the A band dispersion and
for the B band near the gap edge (figure 2(b)). On the other hand,
the MDCs are most useful where the dispersion is steep, e.g. for
the B band between 50 and 100 meV, as well as for the nodal data
of figure 2(a).
\end{references}
\end{document}